\documentclass[%
 reprint,
 amsmath,amssymb,
 aps, pre, superscriptaddress,
]{revtex4-2}

\usepackage{CJKutf8}
\usepackage{graphicx}
\usepackage{dcolumn}
\usepackage{bm}
\usepackage{physics}
\usepackage{color}
\usepackage{hyperref}

\begin{document}
\begin{CJK*}{UTF8}{gbsn}

\title{Non-equilibrium pathways between cluster morphologies in active phase separation: 
necking, rupture and cavitation}

\author{Liheng Yao (姚立衡)}
\affiliation{DAMTP, Centre for Mathematical Sciences, University of Cambridge, Wilberforce Road, Cambridge CB3 0WA, United Kingdom}
\affiliation{Université Grenoble Alpes, CNRS, LIPhy, 38000 Grenoble, France}

\author{Michael E. Cates}
\affiliation{DAMTP, Centre for Mathematical Sciences, University of Cambridge, Wilberforce Road, Cambridge CB3 0WA, United Kingdom}

\author{Robert L. Jack}
\affiliation{DAMTP, Centre for Mathematical Sciences, University of Cambridge, Wilberforce Road, Cambridge CB3 0WA, United Kingdom}
\affiliation{Yusuf Hamied Department of Chemistry, University of Cambridge, Lensfield Road, Cambridge CB2 1EW, United Kingdom}

\begin{abstract}
We investigate the dynamical pathways of a {morphological transition} in a two-dimensional active lattice gas undergoing motility-induced phase separation. The transition is between two {locally stable} morphologies of the liquid cluster: a system-spanning ``slab'' and a compact ``droplet''. We generate trajectories of this transition in both directions using forward flux sampling. We find that the droplet-to-slab transition always follows a similar mechanism to its equilibrium counterpart, but the reverse (slab-to-droplet) transition depends on rare non-equilibrium fluctuations. At low Péclet numbers the equilibrium and non-equilibrium pathways compete, while at high Péclet numbers the equilibrium pathway is entirely suppressed, and the only allowed mechanism involves a large vapour bubble.  We discuss the implications of these findings for active matter systems more generally.
\end{abstract}

\maketitle
\end{CJK*}

\section{Introduction}

Active matter systems are driven away from equilibrium by continuous injection of energy at a microscopic level.  Within this class, systems of self-propelled particles have been widely studied in recent years: they are tractable both analytically and numerically and reveal a range of interesting large-scale behaviour, including motility-induced phase separation (MIPS), where purely repulsive particles spontaneously separate into a dense liquid phase and a dilute vapor~\cite{tailleur2008statistical, fily2012athermal, redner2013structure, cates2025active}.  This phenomenon shares many features with liquid-vapor phase separation at equilibrium, but there are also important differences.  For this reason, MIPS phenomenology is a natural place to test and develop methods for non-equilibrium systems, and to compare with equilibrium cases.

Phase-separated states are also natural settings to investigate the ``controllability'' of many-body physical systems.  For example, the morphology of liquid and vapor domains and the wetting behavior at interfaces can both be engineered via surface tensions between the liquid, vapor, and solid.  It may also be desirable to control the transitions between different {morphologies.  For example, 
a phase-separated system with periodic boundaries may exist in a
 slab morphology, where the liquid domain has flat interfaces and crosses the periodic boundaries; there is  also a droplet morphology with a circular or spherical liquid domain, and curved interfaces.  Both morphologies are locally stable states of phase coexistence.
 In general, the control of  transitions between such states} is intrinsically linked with the mechanism of (often rare) spontaneous transitions via a general connection between optimal-control theory and large deviation theory~\cite{dupuis,Chetrite2015,JackSollich2015,Das2019,Das2021,Heller2024}.  The central idea is that the minimal forces required to induce a dynamical transition are those that realise the  mechanism by which it would happen spontaneously (but rarely).

In fact, characterisation of the transition mechanism between liquid slab and droplet states is already a challenging problem in equilibrium liquid-vapour systems~\cite{venturoli2009kinetics,moritz2017interplay}.  Such rare events are often studied in the context of dewetting and capillary condensation in systems confined between hydrophobic surfaces \cite{lum1998phase, nicolaides1989monte, vishnyakov2003nucleation, luzar2000dynamics, leung2000dynamics, evans1990fluids}.  For the idealised case of the 2d Ising lattice gas with periodic boundaries, it was found by Moritz~\emph{et al.}~\cite{moritz2017interplay} that two reaction coordinates are required to characterise the transition, with one quantifying the neck that forms in the transition state, and the other quantifying the roundness of the cluster.  These co-ordinates have quite different characteristic time scales, so the mechanism involves a relatively slow change of cluster shape, while formation (or breakage) of the neck is much faster.  The surface tension (and its curvature dependence~\cite{troster2011positive, troster2012numerical}) are obviously relevant for this rate, but differing time scales for the two co-ordinates also influence the mechanism, through the presence of two distinct mobilities~\cite{moritz2017interplay}.

This work presents a numerical characterisation of spontaneous transitions between liquid slab and droplet states in an active lattice gas (ALG) that was previously shown to exhibit MIPS~\cite{yao2025interfacial}.  This is achieved by forward flux sampling (FFS), which enables efficient characterisation of such rare events~\cite{Allen2006,Allen2006efficiency,allen2009forward, hussain2020studying}. Compared with the equilibrium case, the situation is more complicated for several reasons.  Firstly, interfacial phenomena in systems undergoing MIPS are significantly different from their equilibrium counterparts, with some effective measures of surface tension giving negative values in certain parameter ranges \cite{bialke2015negative, patch2018curvature, tjhung2018cluster, langford2024theory}.  Moreover, such systems can exhibit bubbly phase separation, where stable bubbles grow within the liquid cluster until they are expelled into the surrounding vapor \cite{tjhung2018cluster, shi2020self, fausti2024statistical, yao2025interfacial}, a phenomenon without any equilibrium equivalent. More exotic interfacial phenomena such as active foam states have also been found \cite{fausti2021capillary}, and there are also interesting analogies with wetting when the liquid phase comes into contact with solid surfaces~\cite{turci2021wetting, turci2024partial, zhao2026wetting, grodzkinski2026hydro, grodzkinski2026spontaneous}.  Understanding and controlling transitions in MIPS systems offers a route towards control of these states.

We study the ALG for parameters such that the slab and droplet morphologies are both locally stable.  For the finite systems that we consider, very long steady-state trajectories would therefore include both slab and droplet morphologies, with rare transitions between them.  (The relative probabilities of the morphologies depends on the particle density, as in the equilibrium case.)  Since the transitions are very rare, it is not efficient to study these transitions by simulating such long trajectories.  Instead, we use FFS to
collect ensembles of trajectories that make transitions between the two morphologies: these are called \emph{reactive trajectories}. Based on these trajectories, we can analyse the  transition mechanism. This application rests on the fact that (i)~FFS is applicable to non-equilibrium systems (the ALG violates detailed balance); and (ii)~FFS does not require knowledge of an accurate reaction coordinate~\cite{allen2009forward}.  These strengths of FFS were already exploited to characterise (rare) nucleation events in systems undergoing MIPS \cite{richard2016nucleation}, and in other  lattice models~\cite{allen2008homogeneous,ryu2010numerical}; additional recent applications of rare event sampling methods in active fluids include~\cite{redner2016classical,das2019variational}.

To briefly summarise our main results, recall that the Péclet number (Pe) quantifies the strength of the particles' self-propulsion.  We consider four state points, two of which have lower values of Pe, and two have higher values.  In all cases, we find that the transition mechanism from droplet to slab resembles that found in~\cite{moritz2017interplay} for the equilibrium case (Ising model).  It is characterised by slow dynamics for droplet shape (``roundness''), both before and after the formation of a neck (which changes the topology of the liquid domain).   We emphasize that the ALG has only repulsive interparticle interactions so its (motility-induced) phase separation is a purely active phenomenon; the passive system of~\cite{moritz2017interplay} is a natural point of comparison but the similarity of droplet-to-slab transitions between the models is already a non-trivial result.
On the other hand, we find that transitions from slab to droplet take place by distinct mechanisms that do not correspond to time-reversed transitions from droplet to slab.  This is a non-equilibrium effect, due to the broken time-reversal symmetry of the steady state.   {We find that the equilibrium-like slab-to-droplet mechanism resembles ductile failure of an amorphous material under mechanical load, including a necking effect~\cite{Hoyle2015,Moriel2018necking,Lin2019distinguishing,Thijssen2023}.  The non-equilibrium mechanisms instead resemble failure by brittle rupture, mediated by voids inside the dense phase and cavitation effects~\cite{Guan2013cavitation,Singh2016,Lin2019distinguishing,Long2021fracture}.  This leads to a qualitative analogy between fracture mechanisms under external stress, and spontaneous transitions in active systems without external loading.

In what follows, we present our model and computational methods in Sec.~\ref{sec:m}, report our findings in Sec.~\ref{sec:r}, and conclude with a discussion in Sec.~\ref{sec:d}.

\section{Model and methods}
\label{sec:m}

\subsection{Model definition}

We use the active lattice gas (ALG) introduced in \cite{yao2025interfacial}, where a 2-dimensional lattice of size $L^2$ and lattice spacing $a$ is populated with $N = \rho (L / a)^2$ particles. Each lattice site can contain at most 1 particle, and the orientation vector of particle $k$ is $\vb{u}_k = (\cos \theta_k, \sin \theta_k)$. This orientation fluctuates over time with a diffusion coefficient $D_O$. Particle $k$ jumps from site $i$ to an unoccupied neighboring site $j$ with rate
\begin{equation}
    W_{ij} = \frac{2 D_E / a^2} {1 + \exp\left[- av_0(\vb{u}_k \cdot \vb{e}_{ij})/D_E \right]}  \; ,
    \label{equ:Wij}
\end{equation}
where $\vb{e}_{ij}$ is a vector that points from site $i$ to site $j$, $D_E$ is the translational diffusion constant, and $v_0$ is the persistence. The rate expression \eqref{equ:Wij} is chosen so that (i) the model converges to that of Mason \emph{et al.}~\cite{mason2023exact} in the limit $a v_0 \ll D_E$, for which an exact hydrodynamic limit has been derived, and (ii) the transition rates satisfy local detailed balance (see e.g. \cite{maes2021local}).

The translational moves are implemented through the Gillespie algorithm \cite{gillespie1977exact, masuda2022gillespie}: the total rate of all allowed transitions $R = \sum W_{ij}$ is computed, and one of the events is picked to happen with probability $W_{ij} / R$, using a binary search tree. The time is then advanced by $1 / R$, and the affected rates $W_{ij}$ are updated.  This process is repeated until the time has been advanced by a total increment $\delta t$, chosen such that $a^2 D_E^{-1} \ll \delta t \ll D_O^{-1}$. After each such increment, the orientation of each particle is updated by an independent Gaussian random variable $\delta \theta_k \sim \mathcal{N} (0, \sqrt{2 D_O \delta t})$, in order to implement the rotational diffusion.

\begin{table}
\centering
\begin{tabular}{| c | c | c | c | c | c |} 
    \hline
    \;State\;  & $v_0$ & $D_E$ & $l_p$ & $\mathrm{Pe}$ & $\rho$  \\ 
    \hline
    A1 & \; $100$ \; & \; $10$ \; & \;$20.0$\; & \;$6.32$\; & \;$0.38$\; \\ 
    \hline
    A2 & $55$ & $50$ & $50.1$ & $7.08$ & $0.45$ \\ 
    \hline
    B1 & $130$ & $30$ & $58.4$ & $10.7$ & $0.38$ \\ 
    \hline
    B2 & $100$ & $50$ & $76.2$ & $10.8$ & $0.41$ \\ 
    \hline
\end{tabular}
\caption{The four state points considered in this work. We have set $a = 1$ and $D_O = 1$ without loss of generality.  {Note that Eq.~\eqref{eqn:pec} expresses $l_p, \mathrm{Pe}$ in terms of $v_0,D_E$ so the state point is fully determined by three parameters which we take as $\mathrm{Pe},l_p$, and the density $\rho$.  The system size is fixed at $L=100$.}}
\label{table:params}
\end{table}

The lengthscales that control the behavior of this model are the diffusive length $l_D = \sqrt{D_E / D_O}$, and the persistence length $l_p$ (see below), which gives the distance traveled by an isolated particle within its orientational diffusion timescale $D_O^{-1}$. The dimensionless Péclet number $\mathrm{Pe} = l_p / l_D$ then measures the strength of the activity in the system, and therefore its distance from equilibrium.  As discussed in~\cite{yao2025interfacial}, the non-linear dependence of $W_{ij}$ on $v_0$ means that $l_p$ and $\mathrm{Pe}$ are given by
\begin{equation}
    l_p = \frac{2 D_E}{a D_O} \tanh \left( \frac{a v_0}{2 D_E} \right) \: , \quad  \mathrm{Pe} = \frac{2}{a} \sqrt{\frac{D_E}{D_O}} \tanh \left( \frac{a v_0}{2 D_E} \right) \: .
    \label{eqn:pec}
\end{equation}
For the purpose of this work, we fix the units of time and space by setting $D_O = 1$ and $a = 1$.  

As shown in \cite{yao2025interfacial}, this model exhibits MIPS and bubbly phase separation for a range of parameter choices. 
We will show results at four state points, which are listed in Table~\ref{table:params}.  Two of these (A1, A2) have relatively low Péclet numbers, but with different diffusive lengths and hence also different persistence lengths; the other two state points (B1, B2) have larger Pe. For all these parameters, the steady state of the system is phase coexistence between a dense liquid cluster and a dilute vapor. The densities are chosen such that the liquid cluster is (meta)stable in both a system-spanning slab geometry and a compact droplet geometry, with rare transitions between these geometries.  All results are for system size $L=100$.  {The spontaneous formation of bubbles in the liquid phase depends significantly on system size~\cite{yao2025interfacial}: for state points B1,B2 and these system sizes, \emph{typical} configurations do not have any bubbles; for A1,A2 some bubbles are visible, see below (Figs.~\ref{fig:A1snaps} and \ref{fig:B2snaps}).}

\subsection{Order parameters}

In modern theories of rare events (see e.g.~\cite{peters2017reaction}), trajectories between {long-lived} states are projected onto one or more order parameters, which are observable quantities that contain information about the mechanism of the transition, and in some cases can be used to calculate transition rates. For transitions with more than one order parameter, a simple signature of irreversibility is that differences appear between forward and backward trajectories when projected onto the order parameters.

Our choice of order parameters follows
Ref.~\cite{moritz2017interplay}, which studied the transition between slab and droplet geometries in the $2d$ Ising model under Kawasaki dynamics \cite{kawasaki1966diffusion}.  The allowed MC moves in that case are particle jumps to nearest unoccupied neighbouring sites, with rates obeying detailed balance with respect to the Ising interaction at temperature $T$.  They used two order parameters to construct a \emph{reaction coordinate} for the transition, which allows the transition rate to be related to a free-energy barrier.

Our first order parameter, denoted $d$, is illustrated schematically in Fig.~\ref{fig:rc}: it measures either the width of a neck that forms in the slab before breaking, or the width of the channel that opens up after breaking. The two situations are differentiated by the sign of $d$, which is positive in the droplet state and negative in the slab state. We note that our definition of $d$ is slightly different from that in \cite{moritz2017interplay}, as we only measure the (integer-valued) distance in the $x$ or $y$ directions, while the authors of \cite{moritz2017interplay} measure the absolute distance (that may not necessarily be in the lattice directions), and add in a constant offset. Our definition makes for easier computational implementation without any significant effect on the physics.

\begin{figure}
    \centering
    \includegraphics[width=0.8\linewidth]{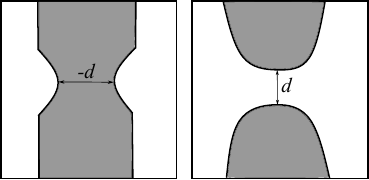}
    \caption{Illustration showing the definition of the reaction coordinate $d$, when the system is in the slab (left) or droplet (right) geometry.}
    \label{fig:rc}
\end{figure}

The second order parameter, denoted $s$, measures the ``roundness'' of the liquid cluster with larger values in the droplet state, and smaller ones in the slab state. Consider the occupancy $\sigma_{xy}$ for the lattice site at position $(x,y)$, which takes the value 1 if the site is occupied and 0 otherwise. We denote the components of its Fourier transform as $\tilde{\sigma}_{pq}$ 
[corresponding to wavevector $2\pi a(p,q)/L$ for integers $p,q$] and calculate
\begin{equation}
    s=\min \left(\frac{\left|\tilde{\sigma}_{01}\right|}{\left|\tilde{\sigma}_{10}\right|}, \frac{\left|\tilde{\sigma}_{10}\right|}{\left|\tilde{\sigma}_{01}\right|}\right) \; .
\end{equation}
Note that a slab oriented along the $y$-axis has large $\tilde{\sigma}_{10}$ and small $\tilde{\sigma}_{01}$; the opposite is true for a slab oriented along the $x$-axis: these both have small $s$.  By contrast the a circular droplet has  $|\tilde{\sigma}_{10}| \approx |\tilde{\sigma}_{01}|$ so $s\approx 1$.

Both coordinates quantify progress along any transition between the two geometries. However, these transitions are controlled by rare events that are impractical to simulate using brute-force methods. Instead we use a rare-event sampling method, as described next.

\subsection{Forward flux sampling}
\label{sec:ffs}

To generate rare trajectories between cluster geometries, we use forward flux sampling (FFS) \cite{allen2009forward}.
Note that this algorithm does not require the system to be in equilibrium; nor does it require accurate knowledge of the reaction coordinate.  {However, sampling efficiency is greatly improved if one does have a good approximation of the reaction co-ordinate~\cite{allen2009forward}, which can make the difference between tractable and intractable calculations.}  The algorithm foliates reactive trajectories along a chosen coordinate, which we choose to be the roundness measure $s$. This choice has various benefits: (a) the value of $s$ varies smoothly between 0 and 1, irrespective of model parameter choices, and (b) $s$ is a slow coordinate compared to $d$, and unlike $d$, does not undergo large fluctuations during the transition.

To perform FFS, we place a set of milestones $s_{i}$ with $i=0,1,2,\dots,m$ in increasing order between 0 and 1. The interfaces are defined such that configurations considered to be in the slab basin have $s < s_{0}$ and those considered to be in the droplet basin have $s \geq s_{m}$.  We describe the method of generating an ensemble of trajectories from the slab state to the droplet.  (The reverse transition is sampled analogously.)  The algorithm depends on a parameter $N_c$, which is the number of reactive trajectories that are sampled. Not all these trajectories are independent, see below.  The method is guaranteed to provide accurate results in the limit of large $N_c$; the convergence of this limit is discussed below.

The first part of the algorithm samples a set of $N_c$ initial conditions for the trajectories.  These have a roundness parameter $s$ between $s_0$ and $s_1$.  They are obtained by preparing a system in the  slab {morphology}, and running a long dynamical trajectory.  We collect configurations whenever $s$ increases from $s < s_0$ to $s \geq s_0$, until we have obtained $N_c$ in total.
The reactive flux $I_{0}$ through the initial interface is given by the number of collected configurations $N_c$ divided by the total duration of this trajectory.

From here, the algorithm proceeds one milestone at a time.  Starting from the ensemble at $i=0$ (which has $s_i<s<s_{i+1}$), we select a random configuration and evolve it until its roundness either reaches $s_{i+1}$ or returns to $s_0$.  This procedure is repeated until $N_c$ configurations have been obtained with $s_i<s<s_{i+1}$: these form the ensemble at milestone $i$.  Increasing $i$ stepwise, we eventually reach the final milestone ($i=m$), at which point $N_c$ reactive trajectories have been obtained.

The probability that a configuration collected at $s_i$ reaches $s_{i+1}$ before it returns to $s_{0}$ can be estimated at each interface, and is denoted $P\!\left(s_{i+1} | s_i\right)$. As configurations with $s > s_{m}$ are considered to be in the product basin with probability $1$, the transition rate is given by
\begin{equation}
    I_{\mathrm{FFS}} = I_{0} \prod_{i=0}^{m-1} P\!\left(s_{i+1} | s_i\right) \; .
    \label{equ:ffsrate}
\end{equation}
In the limit of large $N_c$, this transition rate converges to the physical transition rate.

In our simulations, we place $m = 18$ milestones between $s = 0.08$ and $0.98$, with an even spacing of $0.05$. The placement of the initial interfaces allows crossings of the initial interface to be reasonably spaced apart in time, so that the configurations collected there are not highly correlated. However, our choice also allows crossings to be frequent enough for a large number of configurations to be collected. The spacings between milestones are wide enough that trajectories do not ``overshoot'' (crossing more than one milestone at a time), but narrow enough to still be computationally efficient.

\section{Results}
\label{sec:r}

\begin{figure}
    \centering
    \includegraphics[width=1.0\linewidth]{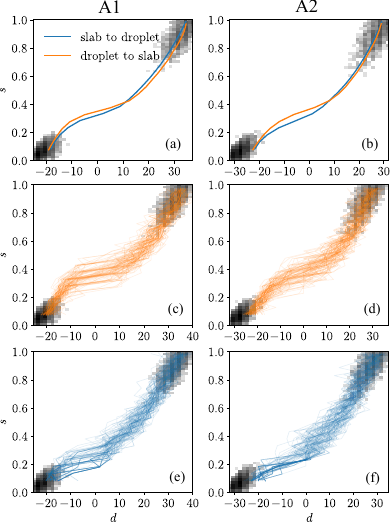}
    \caption{(a,b)~Average transition trajectories of parameters A1 (left column) and  A2 (right column). (c,d) 100 sample trajectories of the droplet to slab transition for each parameter set. (e,f) 100 sample trajectories of the slab to droplet transition. The shadings indicate the probability distributions within the {long-lived slab and droplet states}, obtained by direct sampling. All trajectories are obtained using FFS runs with $N_c = 16000$.}
    \label{fig:lpec}
\end{figure}

\begin{figure}
    \centering
    \includegraphics[width=1.0\linewidth]{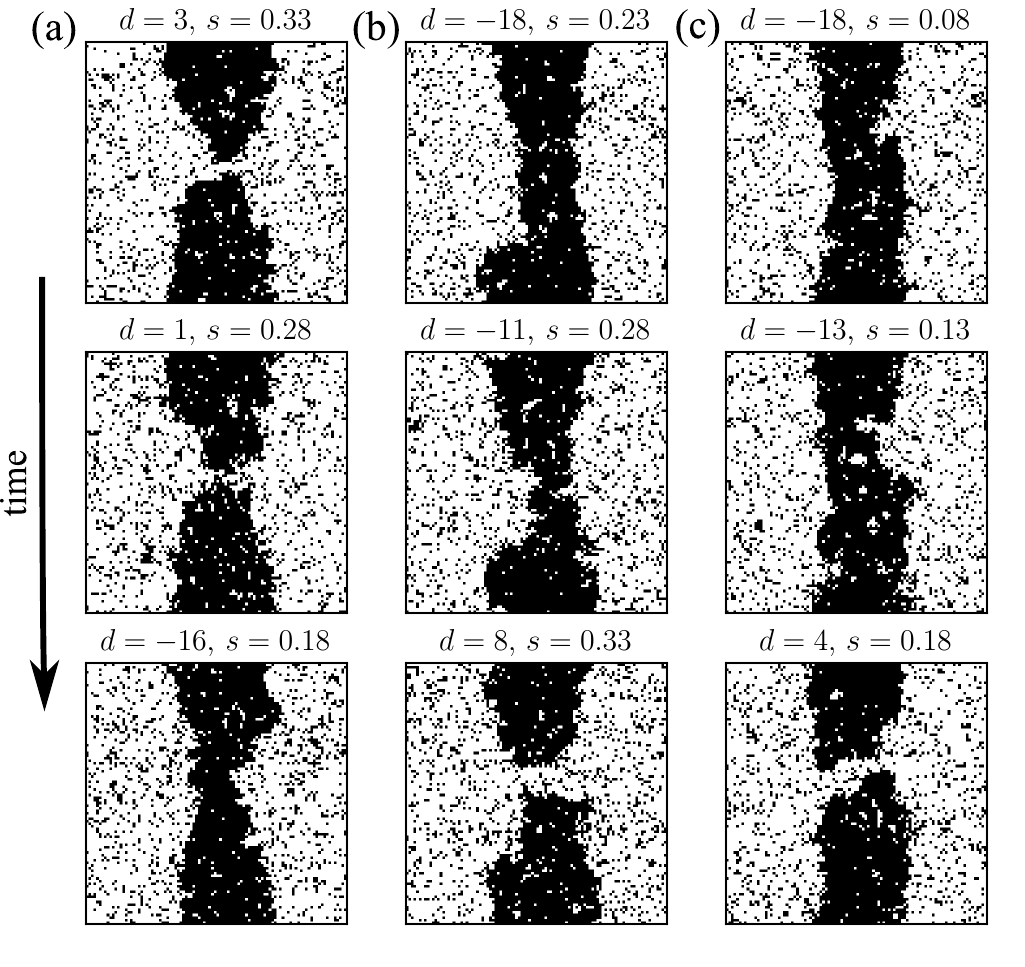}
    \caption{Time-ordered snapshots taken from simulations of parameters A1, showing trajectories of (a) the formation of a slab from a droplet, (b,c) Two trajectories showing the slab-to-droplet transition showing an equilibrium-like mechanism (column b) and a non-equilibrium one (column c).}
    \label{fig:A1snaps}
\end{figure}

We perform FFS for transitions in both the slab-to-droplet and droplet-to-slab directions, for a lattice of size $L = 100$, at densities where both slab and droplet geometries are {locally} stable (recall Tab.~\ref{table:params}). For each configuration collected at the milestones, we measure the values of $s$ and $d$ (the measured $s$ values are all very close to the milestone values, as expected), allowing us to reconstruct reactive trajectories. For each FFS run, we can then average the $s$ and $d$ values at each milestone, yielding the averaged reactive pathway.

We present results for the four state points shown in Tab.~\ref{table:params} which we separate according to their Peclet numbers.  For the smaller Pe (parameters A1,A2) our results in Sec.~\ref{sec:low-pe} are sufficient to establish significant differences between slab-to-droplet and droplet-to-slab transitions.  We find in particular a distinct non-equilibrium mechanism for the slab-to-droplet case, where the slab breaks at a relatively small value of the roundness parameter, which is coupled with density fluctuations (transient bubbles) inside the slab.  We also analyse convergence of the FFS method.  For the larger Pe, the calculations are much more challenging: we present clear numerical evidence in Sec.~\ref{sec:high-pe} for a non-equilibrium slab-to-droplet transition mediated by a large bubble within the slab. 

\subsection{Lower Péclet numbers}
\label{sec:low-pe}

\subsubsection{Pathways}

We first consider parameters A1 and A2 in Table~\ref{table:params}, which  have $\text{Pe}\approx 6.3,7.1$, close to the onset of MIPS, which occurs between $\text{Pe} = 5$ and $\text{Pe}=6$ for the densities we consider.
We show the averaged transition pathways and 100 individual trajectories in Fig.~\ref{fig:lpec}. 
These trajectories are not all independent (see Sec.~\ref{sec:convergence-etc}, below) but they illustrate how the transitions happen in practice, and they indicate the width of the distribution that has been averaged in Fig.~\ref{fig:lpec}.(a,b).}  Some snapshots of the configurations along typical reactive trajectories in Fig.~\ref{fig:A1snaps}. 
For the individual trajectories shown in Fig.~\ref{fig:lpec}(c-f), we note that the branching inherent in FFS means that different trajectories may coincide with each other during the early parts of the transition: these appear as heavier lines in the Figure (we return to this point below). The FFS estimates of transition rates from slab-to-droplet and droplet-to-slab transitions are $3.6 \times 10^{-9}$ and $9.4 \times 10^{-9}$ for parameters A1, and $1.3 \times 10^{-9}$ and $1.3 \times 10^{-9}$ for parameters A2. The ratio of these rates provides an estimate of the relative probabilities of slab and droplet: for parameters A1 the slab state is more stable, while for parameters A2 both states are nearly equally stable.

The droplet-to-slab transition is illustrated in Fig.~\ref{fig:lpec}(c,d) and Fig.~\ref{fig:A1snaps}(a).
We find that
the droplet gradually elongates its shape to leave a narrow channel of vapour; then it  connects quickly across the channel, before relaxing to a stable (``straight'') slab geometry.  The ``fast connections'' appear as the almost-horizontal segments in the individual trajectories shown in Fig.~\ref{fig:lpec}(c,d), showing a sharp changes in $d$ with little change in $s$.
We show time-ordered snapshots of the system just before and after the connection in Fig.~\ref{fig:A1snaps}(c). 
This is the same mechanism that was for found the $2d$ Ising model \cite{moritz2017interplay}.

On the other hand, Fig.~\ref{fig:lpec} shows that the mechanism for slab-to-droplet transitions in the ALG is not the reverse of the droplet-to-slab transition.  This is a non-equilibrium feature: it is different from the Ising case where time-reversal symmetry (detailed balance) means that the averaged trajectories necessarily trace the same path, in opposite directions.
In fact we find considerable diversity in the trajectories from slab to droplet, with two different examples shown in Fig.~\ref{fig:A1snaps}(b,c).  We observe that Fig.~\ref{fig:A1snaps}(b) qualitatively resembles the equilibrium pathway, which is also similar to the time-reversal of Fig.~\ref{fig:A1snaps}(a).
However, inspection of Fig.~\ref{fig:lpec}(e) and (f), shows that this equilibrium-like pathway is in fact outweighed by trajectories where $d$ increases suddenly while $s\lesssim0.2$ is still small, within the range of fluctuations of the slab.

An example of this process can be seen in Fig.~\ref{fig:A1snaps}(c). 
One sees that an indentation forms in the surface of the slab, which then connects with transient bubbles inside the slab, to form a channel of vapor. In equilibrium systems, this dent-like defect is likely to be quickly healed by surface tension, so trajectories involving this configuration are unlikely to carry significant weight in the ensemble of reactive trajectories.
{The existence of such a non-equilibrium transition mechanism is an example of the rich physical behaviour associated with motility-induced phase separation: many features of these systems resemble liquid-vapour coexistence at equilibrium, but signatures of the non-equilibrium behaviour are becoming increasingly clear~\cite{tjhung2018cluster,shi2020self,fausti2021capillary,turci2021wetting,BenDor2022,zhao2026wetting,grodzkinski2026hydro}.  These results may be compared with brittle (rupture) and ductile (necking) behaviour in the breakage of soft materials under external loading, see for example Fig.~5 of~\cite{Lin2019distinguishing}.

\begin{figure}
    \centering	
    \includegraphics[width=0.7\linewidth]{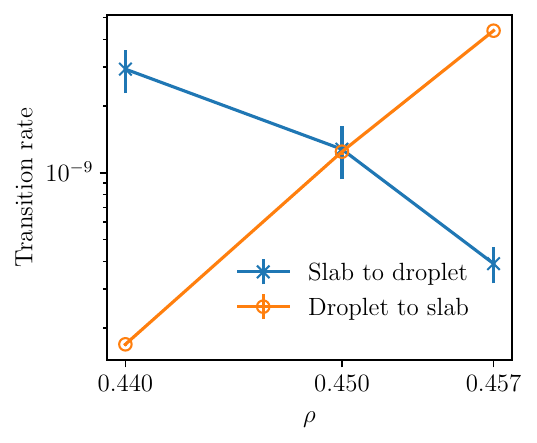}
    \caption{Estimated transition rate as a function of density, for parameters A2. Error bars are standard errors estimated over 3 independent FFS runs at $N_c = 16000$. Note that the error bars for the droplet to slab transition are smaller than the symbol size.}
    \label{fig:rate}
\end{figure}

\subsubsection{Density dependence of the transition rate}
\label{sec:rate}

For parameters A2, we varied the density of the system, to further characterise the transitions. 
Results analogous to those of Fig.~\ref{fig:lpec} are shown in Appendix~\ref{app:sup}.  Here we focus on 
 Fig.~\ref{fig:rate}, which shows the FFS-estimated transition rates as a function of density.  The trends are consistent with physical expectations: increasing density reduces the rate for the slab-to-droplet transition, because the slab is thicker, and is therefore less likely to form a suitable neck or indentation in order to transform to the droplet.  On the other hand, higher density increases the rate of transformation from droplet-to-slab, because the larger droplet requires less deformation in order to cross the periodic boundaries and form a slab.  These are the same trends found in equilibrium systems.

The ratio of slab-to-droplet and droplet-to-slab transitions gives the ratio of the probabilities of the two states, within the steady state. The trends indicate that the slab becomes more probable as the density increases. This effect is also consistent with the equilibrium case, where the interfacial free energy of the slab is $2\gamma L$ (with $\gamma$ being the surface tension) while the interfacial free energy of the droplet is $2\pi \gamma R$ where $R$ is the droplet radius.  On increasing the density at fixed system size $L$, the droplet free energy increases (larger $R$) but the slab remains the same.  Hence the slab becomes more probable, as expected. These results are consistent with the previous result \cite{yao2025interfacial} that this ALG has a positive surface tension in the sense of capillary wave theory.

We also note that for larger densities, a circular liquid droplet cannot fit in the simulation box.  At this point the droplet state no longer exists; one should consider instead a large liquid domain that surrounds a circular domain of the vapour (that is, a macroscopic bubble).  The transition between slab and bubble morphologies would be an interesting direction for future work -- for the Ising case of~\cite{moritz2017interplay} the bubble and droplet morphologies are related by symmetry so the slab-to-bubble transition is analogous to slab-to-droplet, etc.  However, this is not the case in active systems; the asymmetry is particular strong in systems that display bubbly phase separation~\cite{tjhung2018cluster, shi2020self, yao2025interfacial}.


\subsubsection{FFS convergence and physical discussion}
\label{sec:convergence-etc}

\begin{figure}
    \centering
    \includegraphics[width=1.0\linewidth]{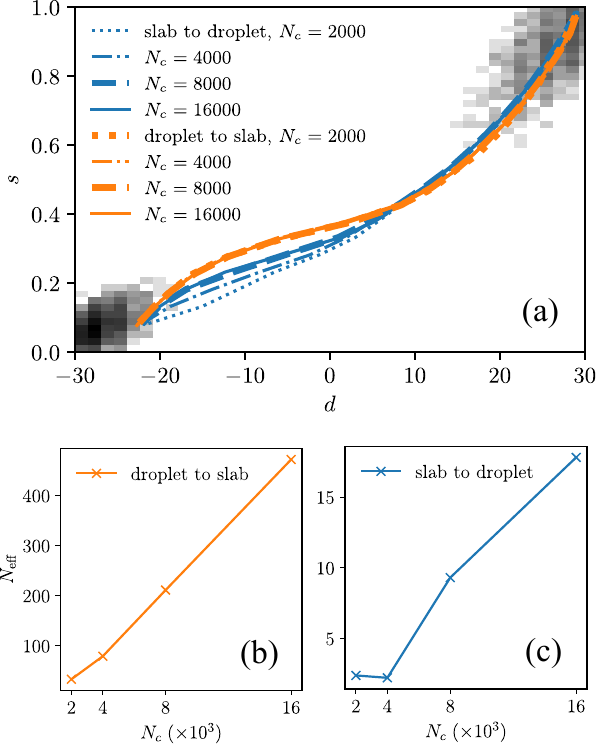}
    \caption{(a) Averaged reactive trajectories for parameters A2 with various values of $N_c$. The shaded densities indicate probability distributions within the {long-lived slab and droplet states}. (b), (c) The effective sample size at the initial milestone $N_{\mathrm{eff}}$ plotted against $N_c$.}
    \label{fig:A2Nc}
\end{figure}

As discussed in Sec.~\ref{sec:ffs}, the FFS algorithm samples representative reactive trajectories if $N_c$ is sufficiently large~\cite{allen2009forward}.  However, convergence of this limit is not simple to achieve in practice~\cite{allen2009forward,vanErp2012}.  In general one expects better performance if the co-ordinate used to define the milestones is a good reaction co-ordinate for the transition of interest.

To investigate convergence, Fig.~\ref{fig:A2Nc}(a) shows averaged reactive trajectories for various $N_c$ for parameters A2.  Each curve comes from a single (representative) run of the FFS algorithm, which yields $N_c$ reactive trajectories.  Then $s$ and $d$ are averaged over the ensembles at each milestone. {For the slab to droplet direction, there are significant run-to-run fluctuations of these averaged trajectories for $N_c \leq 8000$, but the results for the largest $N_c=16000$ are robust.}  This slow convergence indicates that $s$ is not a good reaction co-ordinate in this direction: it does not capture the bubbles that form inside the slab, which are implicated in the non-equilibrium mechanism.

Another way to see this effect is to measure the effective sample size for the reactive trajectories obtained from FFS.  This quantity, defined in Appendix~\ref{app:neff}, measures how many configurations from the first milestone contribute significantly to the sampled ensemble of reactive trajectories. It may be viewed as an approximation to the number of independent initial configurations that contribute to the sampled trajectories. For large $N_c$, {it grows linearly} with $N_c$.

Results are shown in Fig.~\ref{fig:A2Nc}(b,c). The results for the droplet-to-slab transition indicate {excellent FFS performance}, with hundreds of independent trajectories obtained from each run.  However, the effective sample sizes $N_{\rm eff}$ are significantly smaller for the slab-to-droplet transition, which has $N_{\rm eff} <5$ for $N_c \leq 4000$.  This indicates that the sampled trajectories are highly correlated, which explains the significant sample-to-sample fluctuations in this regime.  The relatively poor performance in this case indicates that the  roundness $s$ is not a perfect reaction co-ordinate. In particular, it is not sensitive to small bubbles in the slab, which do affect the transition. {Even so, we stress that even in this case a single run of the FFS algorithm at $N_c = 16000$ yields more than 15 independent trajectories; the results are robust between completely independent runs of the algorithm, as illustrated for example by the sizes of the error bars in Fig.~\ref{fig:rate}.  We recall from Fig.~\ref{fig:rate} that the rate for these rare events is over order $10^{-9}$, each microscopic particle hops between lattice sites with rate $D_E$ between $10$ and $50$, and the systems contain of order $4000$ particles.  Obtaining 10-20 reactive trajectories in such a system is not a simple task; it is a strength of the FFS algorithm that this can be achieved despite $s$ being a far-from-perfect reaction co-ordinate.}

The significant difference in the quality of the equilibrium reaction coordinate $s$ as the FFS order parameter between the two directions is itself a strong indicator of different underlying transition mechanisms: the effectiveness of the co-ordinate $s$ in the droplet-to-slab direction implies an equilibrium-like transition, while poorer performance in the slab-to-droplet direction indicates the existence of other independent transition pathways.  

\begin{figure}
    \centering
    \includegraphics[width=1.0\linewidth]{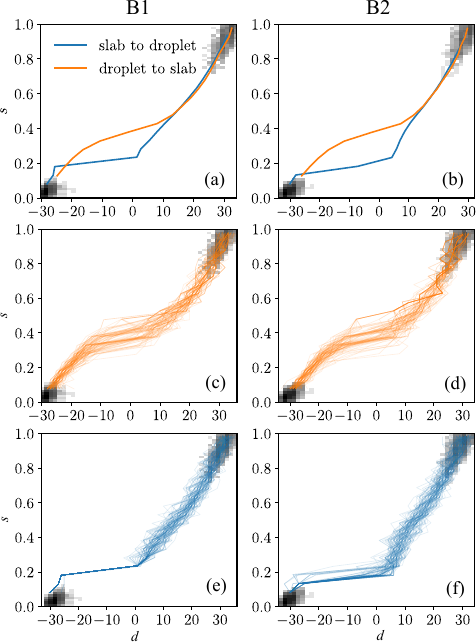}
    \caption{Top to bottom: average transition trajectories of parameters (a) B1 and (b) B2; 100 sample trajectories of the droplet to slab transition for (c) B1 and (d) B2; 100 sample trajectories of the slab to droplet transition for (e) B1 and (f) B2. The density plots are probability distributions within the slab and droplet states. All trajectories are obtained using FFS runs with $N_c = 2000$.}
    \label{fig:hpec}
\end{figure}

\begin{figure}
    \centering
    \includegraphics[width=1.0\linewidth]{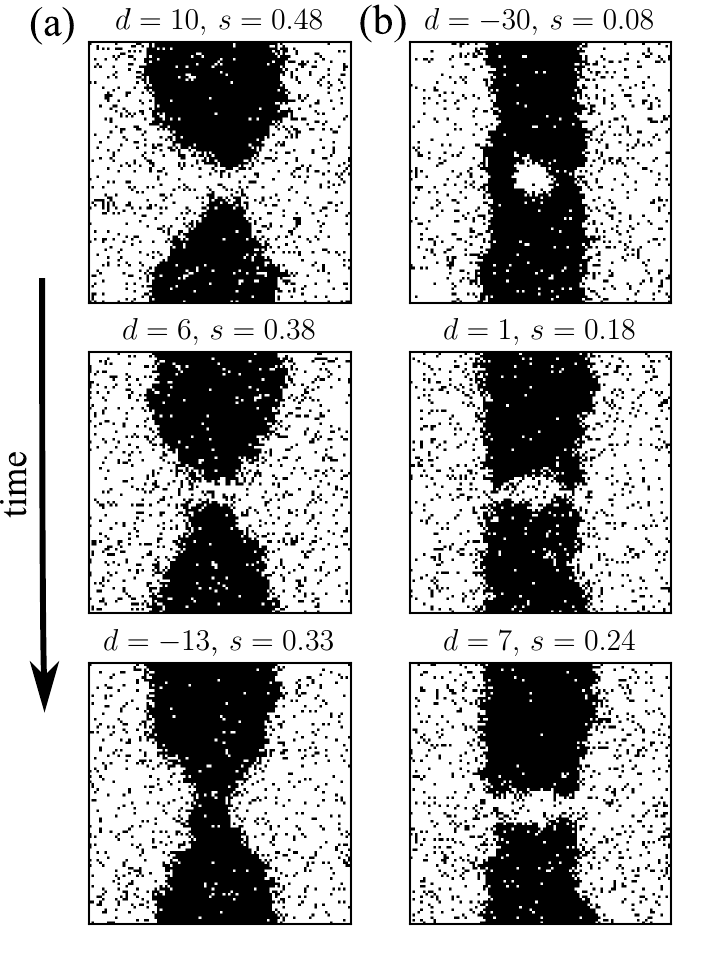}
    \caption{Time-ordered snapshots taken from simulations of parameters B2, showing typical trajectories of (a) the breakage of a slab into a droplet, and (b) the formation of a slab from a droplet.}
    \label{fig:B2snaps}
\end{figure}

\subsection{High Péclet numbers}
\label{sec:high-pe}

We now show results for higher Péclet numbers ($\mathrm{Pe}\approx 10.7$): these are parameters B1 and B2 of Tab.~\ref{table:params}.
The FFS calculations in this case are more computationally expensive compared to the lower Pe (see below for a discussion), so all results of this Subsection have $N_c=2000$.

Again, we show the averaged transition pathways and 100 randomly selected trajectories in Fig.~\ref{fig:hpec}, and some snapshots of the configurations along typical reactive trajectories in Fig.~\ref{fig:B2snaps}. Despite the higher Péclet numbers, the droplet-to-slab transition is again equilibrium-like, as illustrated in the trajectories in Fig.~\ref{fig:hpec}(c,d) and Fig.~\ref{fig:B2snaps}(a). 

However, the slab-to-droplet transition takes place by a different mechanism: for the results of Fig.~\ref{fig:B2snaps}, \emph{all} the collected slab-to-droplet trajectories are observed to contain a large bubble at the initial milestone, from which a rupture develops through the slab.  We show an example of this process in Fig.~\ref{fig:B2snaps}(b), {it is reminiscent of the role of cavitation in the failure of metallic glasses, see for example~\cite{Xi2005fracture,Singh2016}.  In the ALG} at these high Péclet numbers, a flat interface is extremely stable, and shape fluctuations are strongly suppressed. As a result, slab rupture shows up in the $d$-$s$ plane as a sudden increase in $d$ with little change in $s$, as seen in Fig.~\ref{fig:hpec}(e) and (f).  One sees from Fig.~\ref{fig:hpec}(a,b) that the forward and reverse transitions have very different mechanisms.

The non-equilbrium transition mechanism from slab to droplet at high Pe causes poor FFS convergence since the milestones are defined in terms of a poor reaction coordinate~\cite{allen2009forward}.  For each of the FFS calculations from slab-to-droplet in Fig.~\ref{fig:hpec}, all the reactive trajectories originate from just two initial configurations.  That is, bubbles only appear in approximately one in a thousand of the configurations at milestone zero, because such configurations are not selected by the roundness parameter $s$. However, these few configurations will go on to completely dominate the final ensemble of reactive trajectories.  
As a result, at the second and third milestones the algorithm spends a large amount of time firing trajectories from non-bubbly configurations that nearly always return to the initial (slab) state.

This means that the effective sample sizes $N_{\rm eff}$ for both these calculations are between $1$ and $2$: such small numbers indicate that the FFS calculations are not fully converged.  However, the observation of a bubble-mediated transition is reminiscent of the non-equilibrium mechanism found at smaller Pe.  This was observed independently for both parameter sets B1 and B2 and similar results are also shown in Appendix~\ref{app:sup} for an additional case (similar to set B2 but with a reduced density).  This suggests that bubble-mediated transitions are likely to be generic. We also stress that since the FFS co-ordinate $s$ favours equilibrium-like necking, the complete suppression of this mechanism in favour of bubble-mediated rupture strongly suggests that the latter will be the dominant pathway in spontaneous transitions.  We therefore believe that the results of this work provide strong qualitative evidence for a bubble-mediated mechanism, although we have not characterised it quantitatively.

A useful direction for future research would be to develop reaction co-ordinates that are sensitive to bubbles inside the slab, in order to improve the FFS convergence, so that this cavitation mechanism can be characterised quantitatively.  Independently of morphological transitions, the nucleation of bubbles inside MIPS liquids has a wider importance, because of its role in bubbly phase separation~\cite{tjhung2018cluster, shi2020self}.  We are hopeful that future progress on that problem will also facilitate studies of this slab-to-droplet transition (see Sec.~\ref{sec:d} below for further discussion).

\section{Discussion and outlook}
\label{sec:d}

We have presented numerical results for the transition pathways between slab and droplet morphologies in an ALG undergoing MIPS.  The natural comparison is with the work of Moritz~\emph{et al.}~\cite{moritz2017interplay}, which analysed the corresponding transitions in an equilibrium lattice gas with Ising interactions.  We have emphasized that for the active system, the slab-to-droplet and droplet-to-slab transitions generically occur by different mechanisms, because time-reversal symmetry is explicitly broken, via the lack of detailed balance in the underlying dynamics.  

For the droplet-to-slab transition, our results are qualitatively similar to the equilibrium case, as follows. The ``roundness'' co-ordinate $s$ tends to evolve slowly since it requires mass transport over large (hydrodynamic) length scales; starting from a round droplet, $s$ decreases from a value close to $1$ as the droplet becomes increasingly elongated;  the periodic boundaries mean that at some point the rapid formation of a neck becomes favourable and the order parameter $d$ changes sign, signalling a change in topology from droplet to slab.  After this, the roundness $s$ continues to decrease until one reaches a uniform slab with flat interfaces.

As already noted, the slab-to-droplet transition does not generically look like a reversed version of the droplet-to-slab transition. 
For the lower values of Pe considered here (between 6.3 and 7.1), the change of topology from slab to droplet occurs on average at a smaller value of the roundness $s$ [Fig.~\ref{fig:A1snaps}(a,b)].  Fig.~\ref{fig:A1snaps} illustrates how an indentation in an flat slab can combine with vapour domains in its interior, with the result that the slab breaks by a sudden localised process.  After this change -- which occurs at small $s$ -- the resulting liquid domain gradually recovers the circular shape of a droplet ($s\approx 1$).  The results of FFS allow a quantitative characterisation of this transition; the algorithmic efficiency is reduced in this case because the roundness $s$ is less effective as a reaction co-ordinate, but the FFS method still yields many independent reactive trajectories.

For the larger Pe, the slab-to-droplet transition appears different again, in that a vapour bubble nucleates inside the slab (Fig.~\ref{fig:B2snaps}(b)), which triggers the transition to the droplet.   This again leads to a topology change at a relatively small value of $s$, compared to the (equilibrium-like) droplet-to-slab transition.  In this case the roundness $s$ is a poor reaction co-ordinate, which hinders sampling efficiency, so a quantitative analysis of the mechanism is not possible.  However, the qualitative observation of a bubble-nucleation mechanism is robust with respect to parameter changes and consistent with other isntances of spontaneous bubble nucleation int he ALG~\cite{yao2025interfacial} and elsewhere~\cite{tjhung2018cluster, shi2020self}.

There are several possible routes towards a more detailed understanding of these transition pathways in MIPS.  For further numerical work on this model, we expect the most fruitful direction to be development of new order parameters that can capture the reaction coordinates: this would improve the efficiency of the FFS algorithm so that more data would be available, and characterisation of a good reaction co-ordinate would also yield extra insight into the transition.  Other options would be to work on a fluctuating hydrodynamics description of a simlar ALG, similar to~\cite{mason2023exact}, which would allow numerical calculations at continuum level. Alternatively one might work on field-theoretic models such as active model B+~\cite{Wittkowski2014,tjhung2018cluster}.  Such calculations are theoretically challenging but they have the advantage of working with continuous density fields, which avoids the large computational cost of FFS for the ALG, which is built on stochastic simulation of large numbers of microscopic particles.  This would facilitate a more comprehensive survey of parameter space.  Comparison of these different MIPS systems would clarify which of the current results are generic and which are unique to the specific model considered here. 

Finally, we note that the role of vapour ``bubbles'' inside the liquid is consistent with their presence in other aspects of active-matter physics including bubbly phase separation~\cite{tjhung2018cluster, shi2020self, fausti2024statistical}, which has also been observed previously in the typical behaviour of this ALG model~\cite{yao2025interfacial}.   The results of this work emphasize that even if the typical behaviour of phase separated states does not exhibit bubbles, rare events in such systems may still be mediated by bubbly mechanisms.   (We also recall that the observation of bubbles depends significantly on system size in the ALG~\cite{yao2025interfacial} and elsewhere~\cite{shi2020self}, so it would be interesting to explore in more detail the system-size dependence of this mechanism.)  There has been recent progress understanding nucleation in MIPS states~\cite{cates2023classical}; results for that case might help to quantify the probability of the mechanism shown in Fig.~\ref{fig:B2snaps}, and to test whether it also appears in field-theoretic models such as active model B+~\cite{tjhung2018cluster}.

Overall, our results highlight the importance of non-equilibrium fluctuations such as persistent bubbles in morphological transitions in active phase-separated systems. These fluctuations may fundamentally change the mechanisms of the transitions of interest, which are controlled by reaction co-ordinates that are different from the equilibrium case.
Identifying effective non-equilibrium order parameters that can be used to quantify reactive trajectories then becomes important, {because choosing a good order parameter greatly improves the efficiency of rare events sampling algorithms such as FFS.}

\begin{acknowledgments}
RLJ thanks Sophie Bennett for interesting discussions of slab-to-droplet transitions, including some preliminary results for a related model.  We thank the EPSRC for support through grant EP/Z534766/1.
\end{acknowledgments}

\begin{appendix}

\section{Effective sample size}
\label{app:neff}

\begin{figure}
    \centering	
    \includegraphics[width=0.9\linewidth]{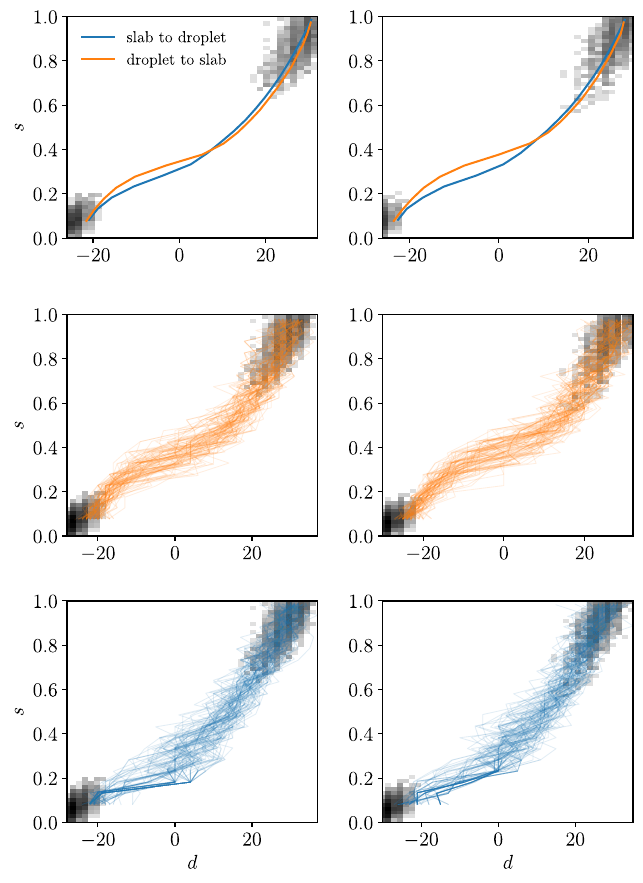}
    \caption{Results analogous to Fig.~\ref{fig:lpec}. The parameters are the same as A2 except that the densities are $\rho = 0.44$ (left) and $\rho = 0.457$ (right).}
    \label{fig:supl}
\end{figure}

The branching nature of the FFS algorithm means that only a subset of the initial conditions (from milestone zero) contributes to the final ensemble of sampled trajectories.  However, sampled trajectories that share the same initial condition are obviously not independent.  (On the other hand, it is expected that trajectories from different initial conditions should be independent, as long as the harvesting of initial conditions is not correlated.) It is useful to quantify the effective sample size associated with this branching, which is achieved by drawing on standard statistical methods~\cite{Liu2008}, as follows.

For the $k$th initial configuration, we write $w_k$ as the number of configurations collected at the final interface that stem from this initial configuration. The effective sample size is then defined as
\begin{equation}
    N_{\rm eff} = \frac{\left( \sum_{k=1}^{N_c} w_k \right) ^2}{ \sum_{k=1}^{N_c} w_k^2 } \; .
\end{equation}
Clearly, if one trajectory is sampled from each initial configuration, then $N_{\rm eff}=N_c$.  On the other hand, if just one initial condition is the source for all trajectories, then $N_{\rm eff}=1$. A larger $N_{\rm eff}$ indicates that more initial conditions are represented in the sampled ensemble of reactive trajectories.

\begin{figure}[t]
    \centering	
    \includegraphics[width=0.5\linewidth]{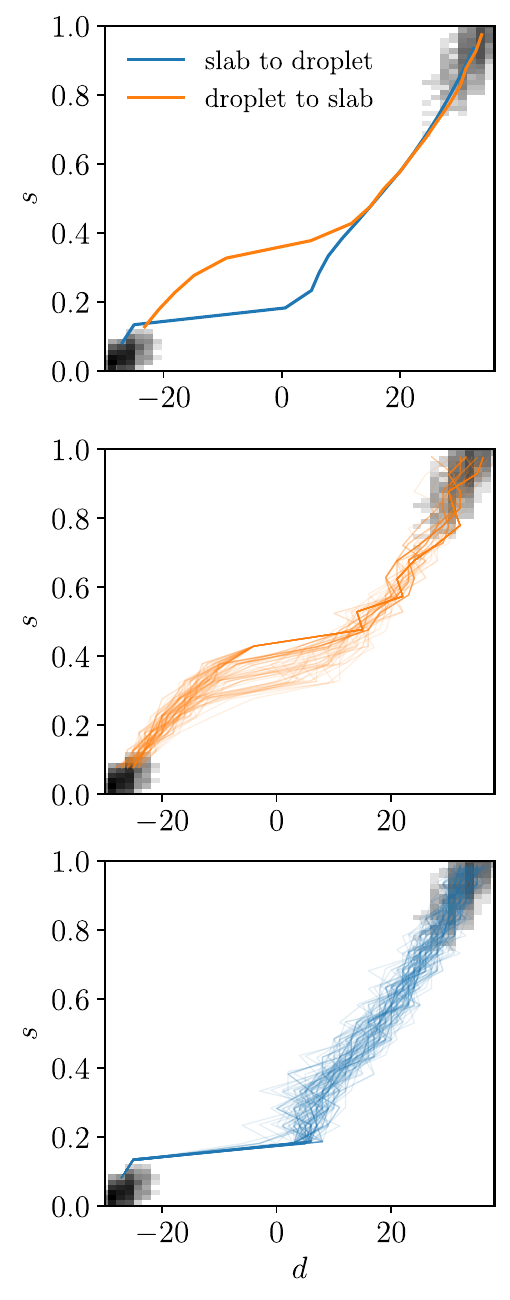}
    \caption{Similar results to Fig.~\ref{fig:hpec} but with a reduced density; the parameters are the same as B2 but with density $\rho=0.38$.}
    \label{fig:suph}
\end{figure}

\section{Density dependence of transition mechanisms }
\label{app:sup}

We present additional data that shows how transition mechanisms change on varying the density $\rho$.  We note in passing that the large interfacial fluctuations of the ALG mean there is a limited range of densities over which both slab and droplet states are locally stable; the purpose of these results is to show that the transition mechanism is robust for modest density changes, within this range.

Fig.~\ref{fig:supl} shows how transition mechanisms change on varying density $\rho$, presented in the same way as Fig.~\ref{fig:lpec} (apart from the density, the parameters are the same as A2).  The corresponding transition rates are shown in Fig.~\ref{fig:rate}.
Changing the density leads to the qualitatively similar behaviour.  The differences between slab-to-droplet and droplet-to-slab transitions becomes more pronounced at high density.  We speculate that this is due the sudden ``rupture'' mechanism of Fig.~\ref{fig:A1snaps}(c) being favoured when the initial slab is thicker, compared to the smooth the ``necking'' mechanism  of Fig.~\ref{fig:A1snaps}(b).

Fig.~\ref{fig:suph} shows the analogous situation at higher Pe (the parameters are the same as B2 but now with a lower density).  We again see that the transition from slab-to-droplet is dominated by a single path where $d$ changes suddenly while $s$ remains almost constant.

\end{appendix}

\bibliography{refs,extra}

\end{document}